\documentstyle{mn}

\newcounter{parentequation}
\def\beglet{
  \addtocounter{equation}{1}%
  \setcounter{parentequation}{\value{equation}}%
  \setcounter{equation}{0}%
  \def\theequation{\arabic{parentequation}\alph{equation}}%
  \ignorespaces
}
\def\endlet{
  \setcounter{equation}{\value{parentequation}}%
  \def\theequation{\arabic{equation}}%
}

\def\ltsima{$\; \buildrel < \over \sim \;$}
\def\gtsima{$\; \buildrel > \over \sim \;$}
\def\simlt{\lower.5ex\hbox{\ltsima}}
\def\simgt{\lower.5ex\hbox{\gtsima}}

\def\etal{{\it et al.}\rm}
\def\etals{{\it et al. }\rm}
\def\mk2{\mu {\rm K}^2}

\begin{document}

\title[Analysis of low CMB Multipoles from WMAP]
{A Maximum Likelihood Analysis of the Low CMB Multipoles from WMAP}

\author[G. Efstathiou]{G. Efstathiou\\
Institute of Astronomy, Madingley Road, Cambridge, CB3 OHA.}

\maketitle

\begin{abstract}
The amplitudes of the quadrupole and octopole measured from the
Wilkinson Microwave Anisotropy Probe (WMAP) appear to be lower than
expected according to the concordance $\Lambda$CDM cosmology. However,
the pseudo-$C_\ell$ estimator used by the WMAP team is non-optimal. In
this paper, we discuss the effects of Galactic cuts on pseudo-$C_\ell$
and quadratic maximum likelihood estimators. An application of a
quadratic maximum likelihood estimator to Galaxy subtracted maps
produced by the WMAP team and Tegmark, de Oliveira-Costa and Hamilton
(2003) shows that the amplitudes of the low multipoles are stable to
different Galactic cuts. In particular, the quadrupole and octopole
amplitudes are found to lie in the ranges $\Delta T_2^2 = 176 - 250\;
(\mu K)^2$ and $\Delta T_3^2 = 794 - 1183\; (\mu K)^2$ (and more
likely to be at the upper ends of these ranges) rather than the values
$\Delta T_2^2 = 123 \; (\mu K)^2$ and $\Delta T_3^2 = 611 \;(\mu K)^2$
found by the WMAP team. These results indicate that the discrepancy
with the concordance $\Lambda$CDM model at low multipoles is not
particularly significant and is in the region of a few percent. This conclusion
is consistent with an analysis of the low amplitude of the angular correlation
function computed from quadratic maximum likelihood power spectrum estimates.

\vskip 0.1 truein

\noindent
{\bf Key words}: cosmic microwave background, cosmology.

\vskip 0.35 truein

\end{abstract}

\section{Introduction}

  Over the last decade or so, a wide range of astronomical data has
suggested that our Universe is described by a `concordance'
$\Lambda$CDM model (see {\it e.g.} Bahcall \etals 1999; Wang \etals
2002).  According to this model, the Universe is spatially flat and
dominated by vacuum energy density and weakly interacting cold dark
matter. In addition, the primordial fluctuations are nearly scale
invariant, as predicted in simple inflationary models of the early
Universe.  The beautiful observations of the cosmic microwave
background (CMB) anisotropies made by the WMAP satellite have added
strong support for this model (Bennett \etals 2003a; Spergel \etals
2003, hereafter S03).

However, the quadrupole and (to a lesser extent) the octopole
amplitudes measured by WMAP are lower than expected according to the
best fitting $\Lambda$CDM model (Bennett \etals 2003a; S03). 
The discrepancy at low multipoles was quantified by S03, who
estimated that the lack of structure at angular scales $\theta > 60^\circ$
on the CMB sky would occur by chance with a probability of only
$1.5 \times 10^{-3}$ if the concordance $\Lambda$CDM model is
correct. This striking result has stimulated a lot of interest,  since it
might indicate the need for exotic new physics (see {\it e.g.}  Efstathiou
2003a; Contaldi \etals 2003; Cline, Crotty and Lesgourgues 2003; Feng
and Zhang 2003; DeDeo, Caldwell and Steinhardt 2003).

However, a number of authors have questioned S03's estimate of the
statistical significance of the discrepancy. Tegmark, de
Oliveira-Costa and Hamilton (2003, hereafter TdOH03) constructed an
all-sky Galaxy subtracted map from the WMAP data and derived higher
amplitudes for the octopole and quadrupole, concluding that the
discrepancy is much less significant (in the region of a few
percent). Efstathiou (2003b, hereafter E03b) argues that errors caused by inaccurate
subtraction of Galactic emission (ignored by S03) should be folded
into the error budget of the low multipoles and that these reduce the
discrepancy to the level of a few percent. Other authors have applied
Bayesian statistics (rather than the frequentist statistics discussed
by S03)  to test whether modified models, {\it e.g.}
 with a sharp break in the power spectrum on large spatial
scales, are preferred to the concordance $\Lambda$CDM cosmology ({\it
e.g.} Bridle \etals 2003; Cline \etals 2003; Contaldi \etals 2003;
Niarchou \etals 2003). Although the Bayesian analyses generally favour
some modification, they do not strongly exclude the simple concordance
$\Lambda$CDM model. A comparison of frequentist and Bayesian
statistics applied to the low CMB multipoles is given by E03b
and will not be discussed further in this paper.

\begin{figure*}

\vskip 5.3 truein

\includegraphics{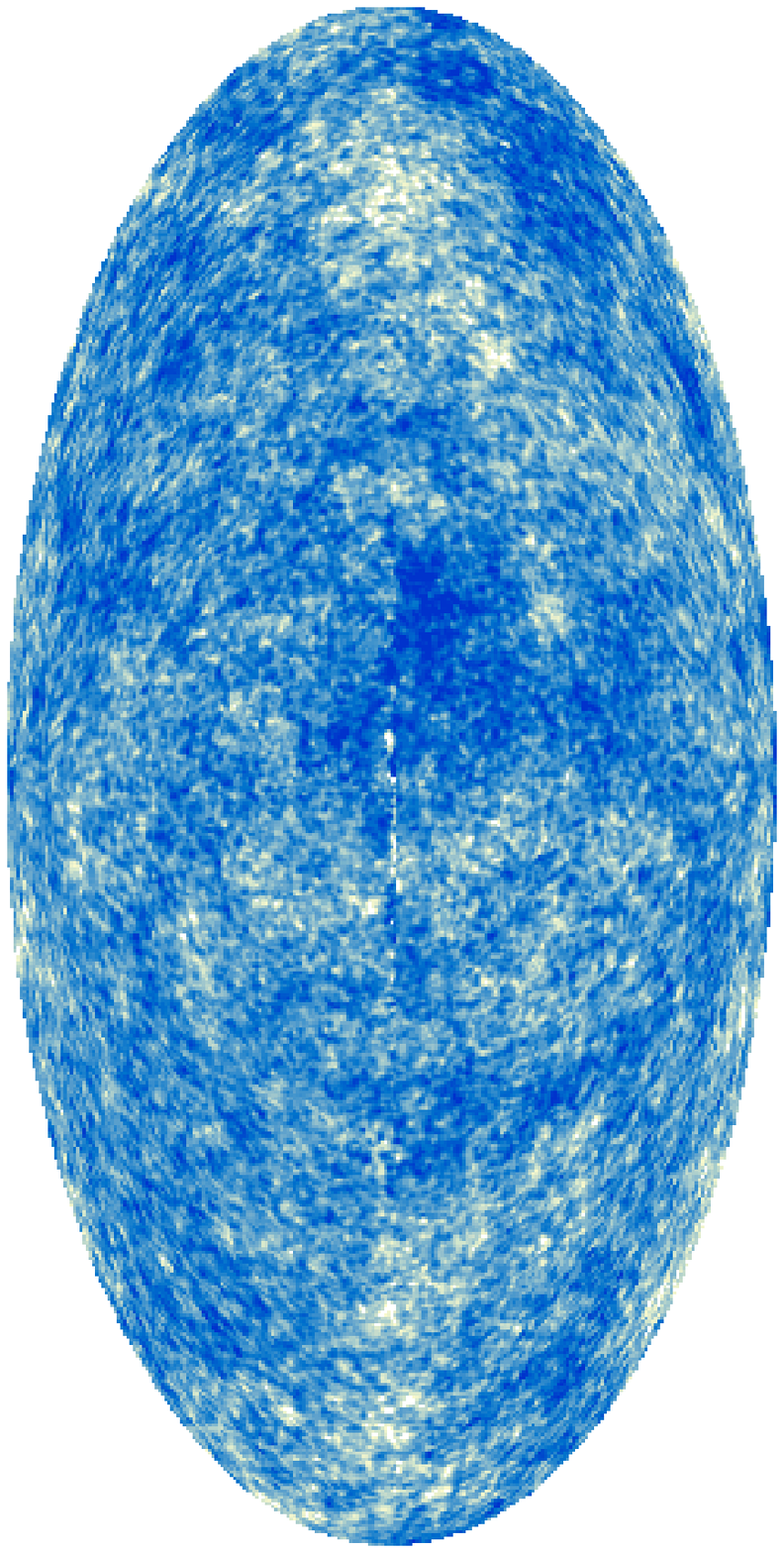}
\includegraphics{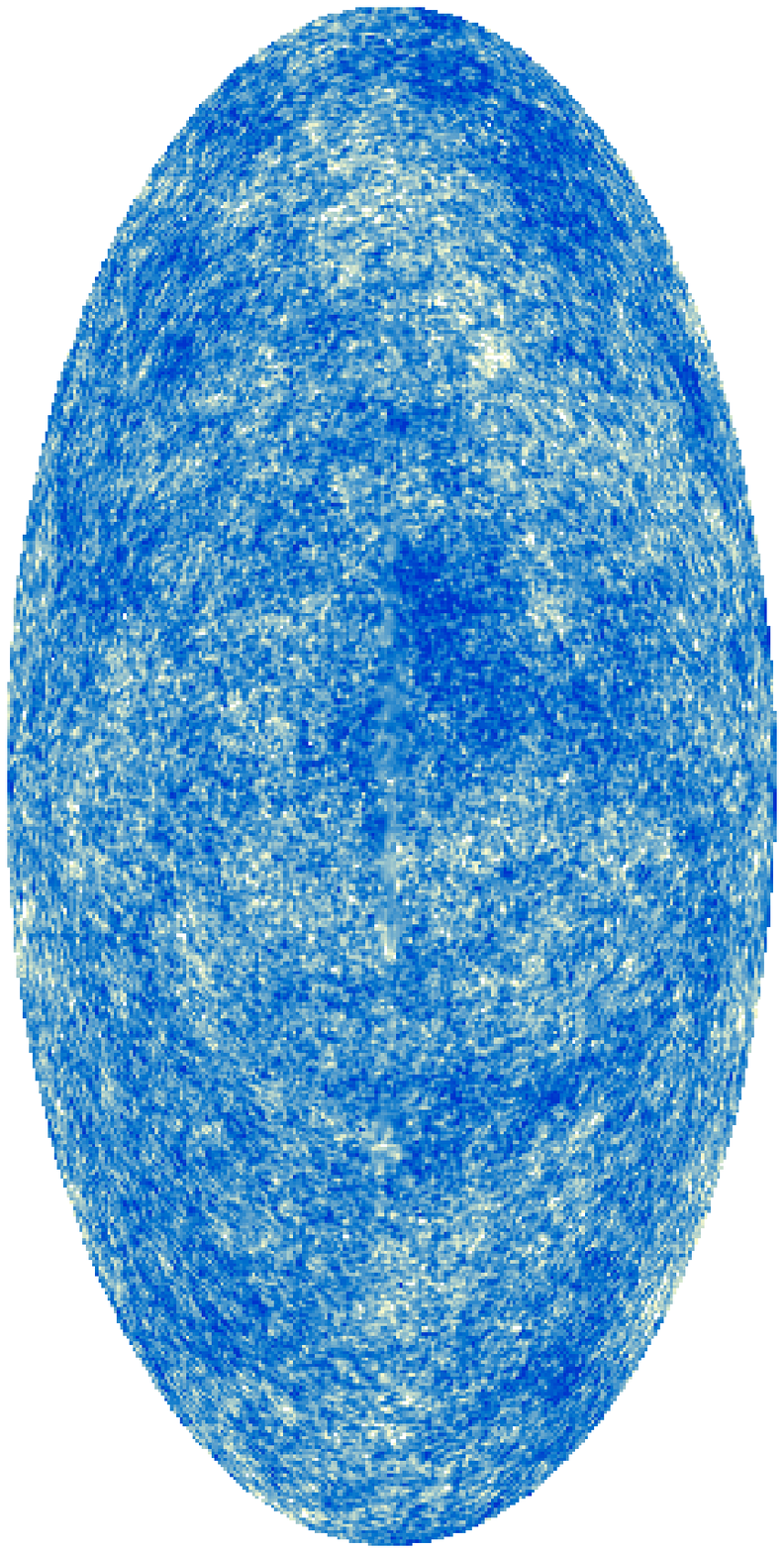}

\caption
{The upper figure shows the WMAP-ILC map of Bennett \etals (2003b)
which is smoothed with a Gaussian beam of $1^\circ$ FWHM. The lower
figure shows the Wiener filtered component separated map of TdOH03. }

\label{figure1}

\end{figure*}

In a previous paper (E03b) I pointed out that the WMAP
analysis of the CMB power spectrum (and angular correlation function)
used a pseudo-$C_\ell$ (hereafter PCL) estimator (Hinshaw \etals
2003). This type of estimator has been discussed extensively in the
literature (see Peebles 1973; Wandelt, Hivon and G\'orski 2001; Hivon
\etals 2002; Efstathiou 2003c, hereafter E03c) and is known to be
non-optimal if applied to an incomplete map of the sky. I also pointed
out that a quadratic maximum likelihood (herafter QML) estimator (see
Tegmark 1997, E03c) can return {\it almost the exact values} for the
low CMB multipoles from an incomplete map of the sky, provided that
the sky cut is not too large and instrumental noise is negligible (a
good approximation for WMAP at large angular scales). It could be
argued, with some justification, that for the relatively modest
Galactic cuts that have been applied to the WMAP data, the type of
estimator used to assess the statistical significance of the low
multipoles should not be particularly critical. For example, if a PCL
estimator is applied with a particular sky cut, the results can be
compared with simulated data using the same estimator and sky cut
(exactly this type of comparison has been done by S03 and E03b).
However, the real CMB sky contains emission from the
Galaxy. As we vary the Galactic cut, how can we tell whether small
changes in the low CMB multipoles are caused by inaccuracies in
Galactic emission or by the effects of the sky cut on the estimator?
This is where a QML estimator can help, since the low CMB multipoles
should be stable to the Galactic cut if a QML estimator is
applied to the data.

As the low CMB multipoles have stimulated so much theoretical
interest, it is surely important to make the most accurate estimates
possible of their amplitudes from the WMAP data, and to test their
sensitivity to residual Galactic emission. This is the aim of this
paper.  

The performance of PCL and QML estimators for different
sky cuts is discussed in Section 2 and tested against numerical
simulations. In Section 3, the QML and PCL estimators are then applied to Galaxy
subtracted maps produced by the WMAP team (Bennett \etals 2003b,
herafter B03b) and by TdOH03 (see Figure 1). Section 4 describes an
analysis of the $S$ statistic  defined by S03 (equation \ref{S1} below).
The conclusions are summarized in Section 5.

\section{Power Spectrum Estimators and Input Maps}

\subsection{Contributions to the Error Budget and Input Maps}

The error budget of an estimator of the CMB power spectrum can be broken down into
four contributions:

\vskip 0.1 truein

\noindent
(i) At low multipoles, the largest source of error is usually
 cosmic variance, which depends on the true form of the
ensemble  average of the CMB power spectrum, $C_\ell$. For full sky
coverage,  the cosmic variance is given by the well known formula
\begin{equation}
  \langle \Delta C_\ell^2 \rangle = {2 C^2_{\ell} \over (2 \ell + 1)}.  \label{I1}
\end{equation}

\vskip 0.1 truein

\noindent
(ii) Next, there is an error associated with the particular estimator
used to evaluate the power spectrum. For example, if a PCL estimator
is used to estimate the power spectrum over a cut sky, the estimates
will differ from the actual values for our particular realization of
sky  since the  cut introduces a loss of information. We will call
this source of error `estimator induced variance'.  It is sometimes
represented, heuristically, by dividing equation (\ref{I1}) by $f_{\rm
sky}$, where $f_{\rm sky}$ is the fraction of the area of the sky that
is unmasked.

\vskip 0.1 truein

\noindent
(iii) Instrumental noise will increase the errors above the cosmic variance
limit of equation (\ref{I1}). For WMAP, instrumental noise is unimportant at
low multipoles and so will be ignored in the rest of this paper.

\vskip 0.1 truein

\noindent
(iv)  Systematic errors from various sources will add to the variance. For the
WMAP data at low multipoles, the most important source of systematic error
is residual contamination from the Galaxy (B03b; Hinshaw \etals 2003).

\vskip 0.1 truein

Cosmic variance is an irreducible component of the error
budget. Accordingly, in this paper we concentrate on estimator induced
variance and systematic errors and ask what are the best estimates of
the amplitudes of the low CMB multipoles for our particular
realisation of the sky?

The effects the Galaxy can be removed by exploiting the frequency
dependence of Galactic emission.  This has been used by B03b to
produce a internal linear combination map (henceforth referred to as
the `WMAP-ILC' map) of the CMB anisotropies, from which the Galaxy has
been subtracted. This map is plotted in the upper panel of Figure 1. A
similar linear combination map, but using the component separation
method of Tegmark and Efstathiou (1996), has been produced by
TdOH03. This map (hereafter referred to as the `TdOH03' map) is
plotted in the lower panel of Figure 1.  The reader is referred to the
papers of B03b and TdOH03 for details of how these maps were produced.

 The low order CMB multipoles can be evaluated from these maps without
imposing a sky cut (TdOH03), in which case the question of estimator
induced variance does not arise since all $4\pi$ steradians of sky are
used. However, it is obvious from visual inspection of Figure 1 that
the Galactic subtraction is not perfect (see also Figure 4), and hence
one might worry about the extent to which the low CMB multipoles are
affected by inaccurate Galactic subtraction. One way of reducing the
effects of Galactic emission still further is by applying a sky
cut. However, as the sky cut is made larger, the estimator induced
variance will generally increase. The main goal of this paper is to
investigate the effects of sky cuts on the low CMB multipoles using an
estimator for which the estimator induced variance is demonstrably
small.

\subsection{PCL estimator}

The PCL estimator is constructed from the spherical harmonic transform of the
map
\begin{equation}
 \tilde a_{\ell m} = \sum_i w_i \Delta T_i \Omega_i Y_{\ell m}({\bf \theta}_i),    \label{PCL1}
\end{equation}
where $\Omega_i$ is the solid angle of pixel $i$ and $w_i$ is a pixel
weight function which henceforth will be set to unity for pixels
outside the sky cut and to zero  within the cut . From these
spherical harmonic coefficients we can construct an unbiased estimator
of the power spectrum,
\begin{equation}
 \hat C^p_\ell = M^{-1}_{\ell \ell^\prime} \tilde C^p_{\ell^\prime}, 
\quad
   \tilde C^p_\ell = {1 \over (2 \ell + 1)} \sum_m \vert \tilde a_{\ell m} \vert ^2,  \label{PCL2}
\end{equation}
where $M$ is a coupling matrix which can be expressed in terms of the power spectrum
of the weight function $w_i$ and 3-j coefficients (see Hivon \etals 2001). Analytic
expressions for the covariance matrix of the PCL estimator accurate at
high multipoles are given by Hinshaw \etals (2003), Chon \etals (2003) and E03c,
and at low multipoles by E03c.

 As is well known, the PCL estimator is sub-optimal and the estimator
induced variance increases as the sky cut is increased, though for
modest sky cuts (removing,  say,  15 \% of the sky), the estimator induced
variance is small in comparison to the cosmic variance of 
the concordance $\Lambda$CDM cosmology (E03c). The
analysis of the low CMB multipoles discussed by Hinshaw \etals (2003)
uses the PCL estimator, $\hat C^p_\ell$, of equation (\ref{PCL2}).

\subsection{QML estimator}

The QML estimator (Tegmark 1997) in the limit of negligible instrumental noise
is
\beglet
\begin{equation}
 y_{\ell} = x_i x_j E^\ell_{ij}, \label{ML1a}
\end{equation}
where $x_i$ is the data vector. (In the application 
discussed in this paper, the data vector consists of the 
temperate differences $\Delta T_i$ over the unmasked sky).
The matrix $E^\ell$ in equation (\ref{ML1a}) is
\begin{equation}
 E^\ell = {1 \over 2}C^{-1} {\partial C \over \partial C_\ell} C^{-1}, \label{ML1b}
\end{equation}
\endlet 
where $C$ is the covariance matrix $\langle x_i x_j
\rangle$. The covariance matrix of the estimates $y_\ell$ is given by
\begin{equation}
F_{\ell \ell^\prime} = \langle y_\ell y_{\ell^\prime} \rangle 
- \langle y_\ell \rangle \langle y_{\ell^\prime} \rangle 
= 2 \; {\rm Tr} \left [ C E^{\ell} 
C E^{\ell^\prime} \right ],  \label{ML3} 
\end{equation}
(Tegmark 1997).

From the estimates (\ref{ML1a}), we can form an unbiased estimate
of the CMB power spectrum
\begin{equation}
  \hat C^q_{\ell}  = F^{-1}_{\ell \ell^\prime} y_{\ell^\prime}
, \label{ML4}
\end{equation}
with covariance matrix
\begin{equation}
 \langle \Delta \hat C^q_{\ell} \Delta \hat C^q_{\ell^\prime}
\rangle  = F^{-1}_{\ell \ell^\prime}.  \label{ML5}
\end{equation}

\begin{table*}
\bigskip

\centerline{\bf \ \ \  Table 1:  
Estimates of Estimator Induced Variance  from Simulations}
\begin{center}

\begin{tabular}{c|cccccc|cccccc} \hline \hline
\smallskip 
 & \multicolumn{6}{c} {QML}&  \multicolumn{6}{c} {PCL} \cr
$\Delta T^2_\ell$ & \multicolumn{2}{c} {all}  & $<1000$ & $<500$ & $<250$ & $<1000$ &
\multicolumn{2}{c} {all}  &  $<1000$ & $<500$ & $<250$ & $<1000$ \cr
 & $\ell = 2$ & $\ell = 3$ & $\ell=2$ & $\ell=2$& $\ell=2$& $\ell=3$&
 $\ell = 2$ & $\ell = 3$ &  $\ell=2$ & $\ell=2$ & $\ell=2$ & $\ell=3$ \cr
Kp2 & $45$ & $50$ & $32$ & $26$ & $18$ & $41$ & $244$ & $203$ & $134$ & $84$  & $55$ & $132$\cr
Kp0 & $95$ & $91$ & $65$ & $51$ & $37$ & $71$ & $355$ & $283$ & $190$ & $116$ & $75$ & 
$181$\cr
\hline
\end{tabular}
\smallskip

\begin{quote}
{\it Notes to Table 1:} The entries list the {\it rms} differences (in
$(\mu K)^2$) between the estimated and input quadrupole and octopole
amplitudes determined from $5000$ simulations with the WMAP Kp2 and
Kp0 sky masks. The columns marked `all' show the results from all $5000$
simulations. The remaining columns list the results with various restrictions
placed on the amplitudes of the quadrupole and octopole in the simulations
(see text for details).
\end{quote}

\end{center}

\smallskip

\end{table*}

The behaviour of the estimator (\ref{ML4}) in the limit of low multipoles
has been discussed by E03c. Let the data vector $x_i$ consist of
the harmonic coefficients $\tilde a_{\ell m}$ defined in equation (\ref{PCL1}).
These coefficients are related to the true $a_{\ell m}$
coefficients on the uncut sky by a coupling matrix $K$,
\begin{equation}
 \tilde a_{\ell m} = \sum_{\ell^\prime m^\prime} a_{\ell^\prime m^\prime}
K_{{\ell m}{\ell^\prime m^\prime}}. \label{ML6}
\end{equation}
If some of the sky is removed by a sky cut, the matrix $K$ is equation
(\ref{ML6}) will be singular. This simply expresses the fact that
there is no information on anisotropies that lie within the sky cut,
hence it is impossible to reconstruct all of the $a_{\ell m}$ from
the harmonic coefficients $\tilde a_{\ell m}$ measured on the cut
sky. For small sky cuts and low multipoles, however, it may be a good
approximation simply to truncate the summation in (\ref{ML6}) at
finite values of $\ell^\prime$ and $m^\prime$. The true low multipole
coefficients $a_{\ell m}$ can then be reconstructed by inverting the
non-singular truncated matrix $\tilde K$. In this case, the QML estimator
becomes 
\begin{equation}
\quad
   \hat C^q_\ell \approx  {1 \over (2 \ell + 1)} \sum_m \vert  a_{\ell m} \vert ^2,  
\qquad a = \tilde K^{-1} \tilde a, 
\label{ML7}
\end{equation}
and is independent of the assumed form for the true power spectrum .
(The variance on these estimates, is given by the cosmic variance 
of equation (\ref{I1}) and does, of course, depend on the assumed 
form for the input power spectrum).

The implication of the above analysis is that for small sky cuts, it
is possible to reconstruct the exact power spectrum coefficients for
our particular realisation of the sky using a QML estimator.  In other
words,  estimator induced variance can be reduced to negligible
levels if a QML estimator is used to estimate the low multipoles,
provided the sky cut is small enough.

\subsection{Simulations of Estimator Induced Variance}

\begin{figure*}
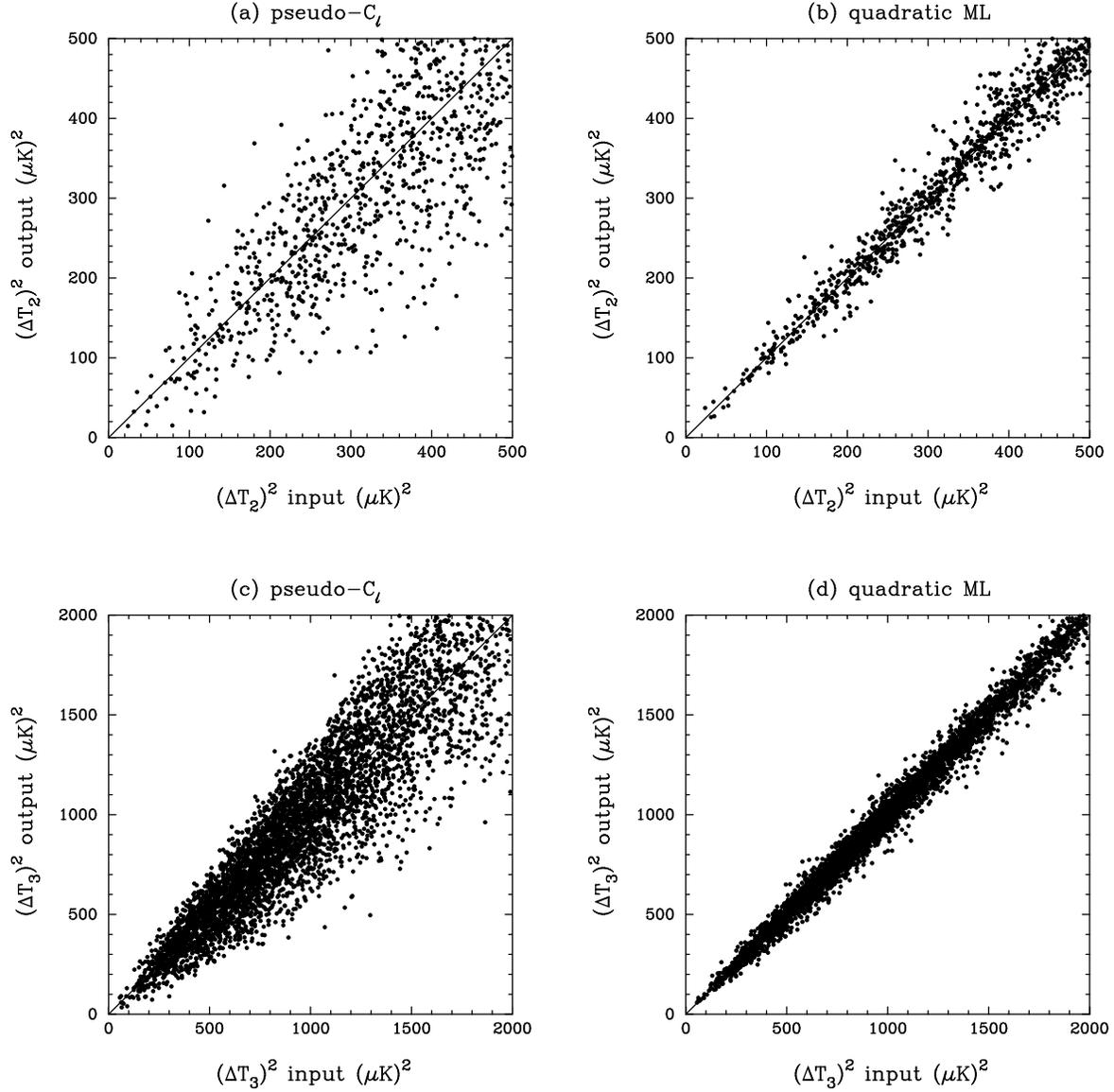


\vskip 6.5 truein

\includegraphics{pgc2a_kp2.ps}
\includegraphics{pgc2b_kp2.ps}
\includegraphics{pgc2c_kp2.ps}
\includegraphics{pgc2d_kp2.ps}

\caption
{Comparison of the PCL and QML estimates of the quadrupole and
octopole from simulations with the Kp2 mask. The abscissae list the
input values of the quadrupole and octopole used to generate the 
simulated skies. The ordinates give the output values from the
PCL estimator (Figures 2a and 2c) and QML estimator (Figures 2b and
2d) after the application of the Kp2 mask.}

\label{figure2}

\end{figure*}

\begin{figure*}
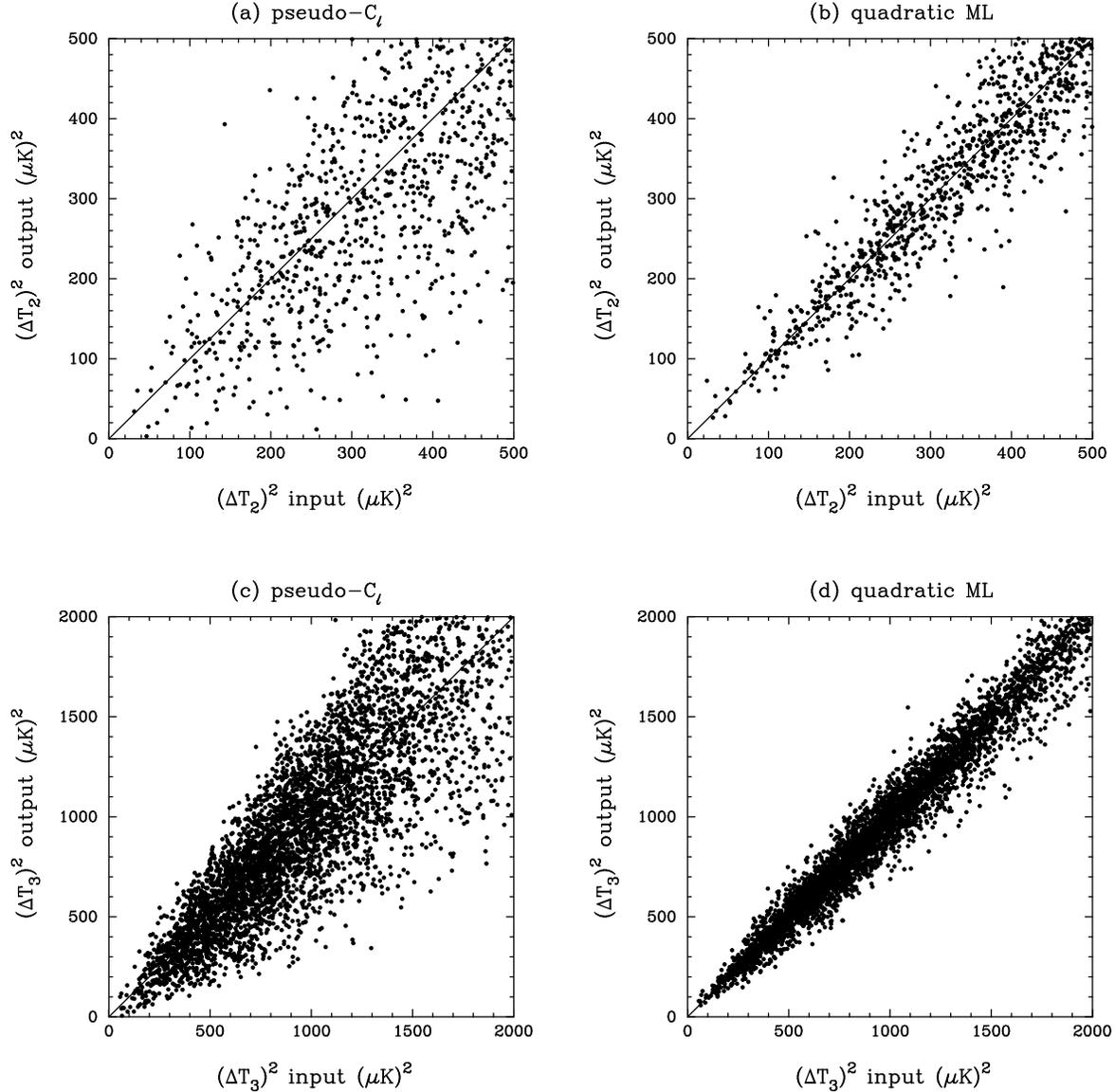


\vskip 6.5 truein

\includegraphics{pgc2a.ps}
\includegraphics{pgc2b.ps}
\includegraphics{pgc2c.ps}
\includegraphics{pgc2d.ps}

\caption
{As Figure 2, but for simulations with the Kp0 mask.}

\label{figure3}

\end{figure*}

Equation (\ref{ML7}) is approximate and will not apply for a large sky
cut. How large does the sky cut have to be for equation (\ref{ML7}) to
break down? As a rough rule of thumb, equation (\ref{ML7}) will be
approximately correct if the the angular correlation function
$C(\theta)$ can be estimated on the cut sky on all angular scales
(Mortlock, Challinor and Hobson 2002). However, for a more precise
assessment, the results of several sets of numerical simulations are
describe in this sub-section. (Similar simulations are described in
EO3c).

 A set of $5000$ noise-free Gaussian CMB maps were generated using the
igloo pixelization scheme described in E03c. The input CMB power
spectrum for these simulations is that of a simple six parameter
$\Lambda$CDM model that provides an excellent fit to the WMAP power
spectrum. This model will be referred to as the `fiducial
$\Lambda$CDM model' in this paper. The values of the six parameters of
are given in E03b. A pixel size of $\theta_c = 5^\circ$ was used and
the maps smoothed with a Gaussian beam of FWHM $7^\circ$.  The WMAP
Kp2 and Kp0 masks (see B03b for a discussion of the WMAP masks) were
repixelised from the HEALPIX NSIDE=512 format (see G\'orski \etals
1999) onto the $\theta_c = 5^\circ$ igloo pixelization.  The PCL and
QML estimators, as described in the preceding sections, were then
applied to each of the simulated maps after masking first by the Kp2
mask (which removes $17\%$ of the sky at this resolution, see Figure
4b) and then by the Kp0 mask (which removes $25\%$ of the sky, see
Figure 4c).

Table 1 lists the {\it rms} differences between the estimated octopole
and quadrupole amplitudes\footnote{Throughout this paper the amplitudes of the CMB multipoles will be expressed
as 
$$
   \Delta T_\ell^2 =  { 1 \over 2 \pi}  \ell (\ell + 1) C_{\ell}.
$$
}, $\Delta T^2_\ell$ and the true amplitudes, $(\Delta T^2_\ell)^T$,
 for the QML and PCL estimators.  The numbers under the columns
 marked `all' show the results for all $5000$ simulations. For the Kp0
 mask, the estimator induced variance of the PCL estimator at low
 multipoles is comparable to the cosmic variance for the fiducial
 $\Lambda$CDM model ($721\; (\mu K)^2$ for the quadrupole and $567\;
 (\mu K)^2$ for the octopole). Evidently, the Kp0 mask is close to the
 transition point at which the PCL estimator breaks down. The
 remaining entries in the table list the {\it rms} differences with
 various restrictions on the input quadrupole and octopole amplitudes
 of the simulations. Thus, Table 1 lists the differences if the input
 quadrupole amplitude is restricted to be less than $1000 \; (\mu
 K)^2$, $500 \; (\mu K)^2$ and $250 \; (\mu K)^2$, and if the octopole
 amplitude is restricted to be less than $1000 \;(\mu K)^2$. In all
 cases, estimator induced variance for the PCL estimator is several
 times larger than for the QML estimator.  The results with the
 restrictions $(\Delta T^2_2)^T < 250 \; (\mu K)^2$ and $(\Delta T^2_3)^T <
 1000 \;(\mu K)^2$ should give reasonable indications of the
 estimator induced variance, since these limits are close to the actual
 measured values of the octopole and quadrupole amplitudes (see
 Section 3).

 The {\it rms} differences listed in Table 1 give a misleadingly
favourable impression of the performance of the PCL estimator because
the distribution of the differences is non-Gaussian. This is illustrated
by Figures 2 and 3 which show the estimated octopole and quadrupole
amplitudes plotted against the input amplitudes for both
estimators. Figure 2 shows the results for the simulations with the
Kp2 mask and Figure 3 shows the results for the Kp0 mask. In
particular, Figure 3a, shows how poorly the PCL estimator behaves when
applied to simulations with the Kp0 mask.  Here there are several
examples of simulations with estimated quadrupole amplitudes of
$\Delta T^2_2 \approx 100\; (\mu K)^2$ where the true quadrupole amplitude is
greater than $300 \; (\mu K)^2$. In contrast, the estimator induced
variance for the QML is much smaller and more closely Gaussian
distributed with a dispersion of $\approx 40 \; (\mu K)^2$). The
estimator induced variance for the QML estimator is almost a factor of
two smaller in the simulations with the Kp2 mask and for most
purposes can be ignored.

 The results of this Section have quantified, by direct numerical
simulation, the effects of estimator induced variance on QML and PCL
estimators. As expected, when applied to maps with identical sky cuts the
QML estimator gives much more accurate estimates of the low multipoles
than the PCL estimator. For the Kp0 mask, which removes $25\%$ of the
sky, the estimator induced dispersion of the quadrupole amplitude is
comparable to the actual measured value of the quadrupole.  If one
applied a mask that removed, contiguously, an even larger number of
pixels than the Kp0 mask, the PCL estimator induced dispersion of the
quadrupole amplitude would become larger than the signal. In contrast,
the estimator induced variance for the QML estimator is negligible for
the Kp2 mask, and contributes a dispersion of about $40\; (\mu K)^2$ for
the quadrupole if the Kp0 mask is imposed.

Before applying the QML and PCL estimators to the WMAP component
separated maps of Figure 1, it is worth anticipating some of the
results:

\vskip 0.1 truein

\noindent
{\it (i) Analysis of all sky component separated maps:} If  the
PCL and QML estimators are applied to all sky maps, the two estimators should give
identical results because in this limit they are mathematically
equivalent. Any differences between the estimated power spectrum and
the primordial power spectrum for our realisation of the sky must then be
caused by systematic errors. The differences between the power spectra
estimated from the WMAP-ILC and TdOH03 maps would provide some
indication of the effects of inaccurate subtraction of Galactic
emission. 

\vskip 0.1 truein

\noindent
{\it (ii) Analysis of component separated maps with the Kp2 mask:} The
estimator induced variance for the QML estimator is negligible in this
case, hence any differences with the results from the all sky maps, or
between the WMAP-ILC and TdOH03 maps, are likely to be caused by
inaccurate subtraction of Galactic emission.  Differences between the
QML and PCL estimators will indicate the effects of estimator induced
variance in the PCL estimates.

\vskip 0.1 truein

\noindent
{\it (iii) Analysis of component separated maps with the Kp0 mask:}
Small differences between the QML estimates applied to maps with the
Kp0 mask and Kp2 masks could be caused by residual Galactic
contamination or by estimator induced variance. Any differences will be
consistent with estimator induced variance if they are of order $40\;
(\mu K)^2$ for the quadrupole and of order $100\; (\mu K)^2$ for the
octopole. With the Kp0 mask, it would not be suprising to find large
difference between the PCL and QML estimates, since the estimator
induced variance for the PCL estimates is expected to be comparable to
the measured amplitudes.

\begin{table*}
\bigskip

\centerline{\bf \ \ \  Table 2:  
Amplitudes of the  Quadrupole and Octopole from WMAP}

\begin{center}

\begin{tabular}{ccc|cc|cc|cc|cc} \hline \hline
\smallskip 
 & &  &\multicolumn{4}{c|}{WMAP-ILC}& \multicolumn{4} {c|}{TdOH03} \cr
 & & & \multicolumn{2}{c} {QML}&  \multicolumn{2}{c} {PCL} & \multicolumn{2}{c}{QML} & 
\multicolumn{2}{c} {PCL} \smallskip \cr
mask & $f_{\rm sky}$ & $\sigma_{\rm maps}$ & $\Delta T^2_2$ & $\Delta T^2_3$ & $\Delta T^2_2$& $\Delta T^2_3$&
 $\Delta T^2_2$ & $\Delta T^2_3$ &  $\Delta T^2_2$ & $\Delta T^2_3$ 
\smallskip \cr
0 &  $1.00$ & $8.9$  & $192$ & $1039$ & $194$ & $1052$ & $198$ & $852$ & $201$ & $865$ \cr
Kp2$\;\;$ & $0.83$ & $6.2$ &  $223$ & $1183$ & $116$ & $480$ & $250$ & $1081$ & $154$ & $471$ \cr
Kp0$\;\;$ &  $0.75$ & $5.6$ & $176$ & $805$ & $ 97$ & $325$ & $245$ & $794$ & $142$ & $331$ \cr
Kp0+ & $0.64$ & $4.1$&  $182$ & $817$ & $120$ & $289$ & $231$ & $804$ & $145$ & $292$ \cr
\hline
\end{tabular}

\smallskip
\begin{quote}
{\it Notes to Table 2:} The second column lists the fraction of sky
that is unmasked.  The third column (labelled $\sigma_{\rm maps}$)
lists the {\it rms} temperature difference in $\mu K$ of the WMAP-ILC
and TdOH03 maps in the unmasked region of the sky. The remaining
columns list the quadrupole and octopole amplitudes in $(\mu K)^2$
determined from the QML and PCL estimators.
\end{quote}

\smallskip

\end{center}

\end{table*}

\begin{figure*}

\vskip 5.0 truein

\includegraphics{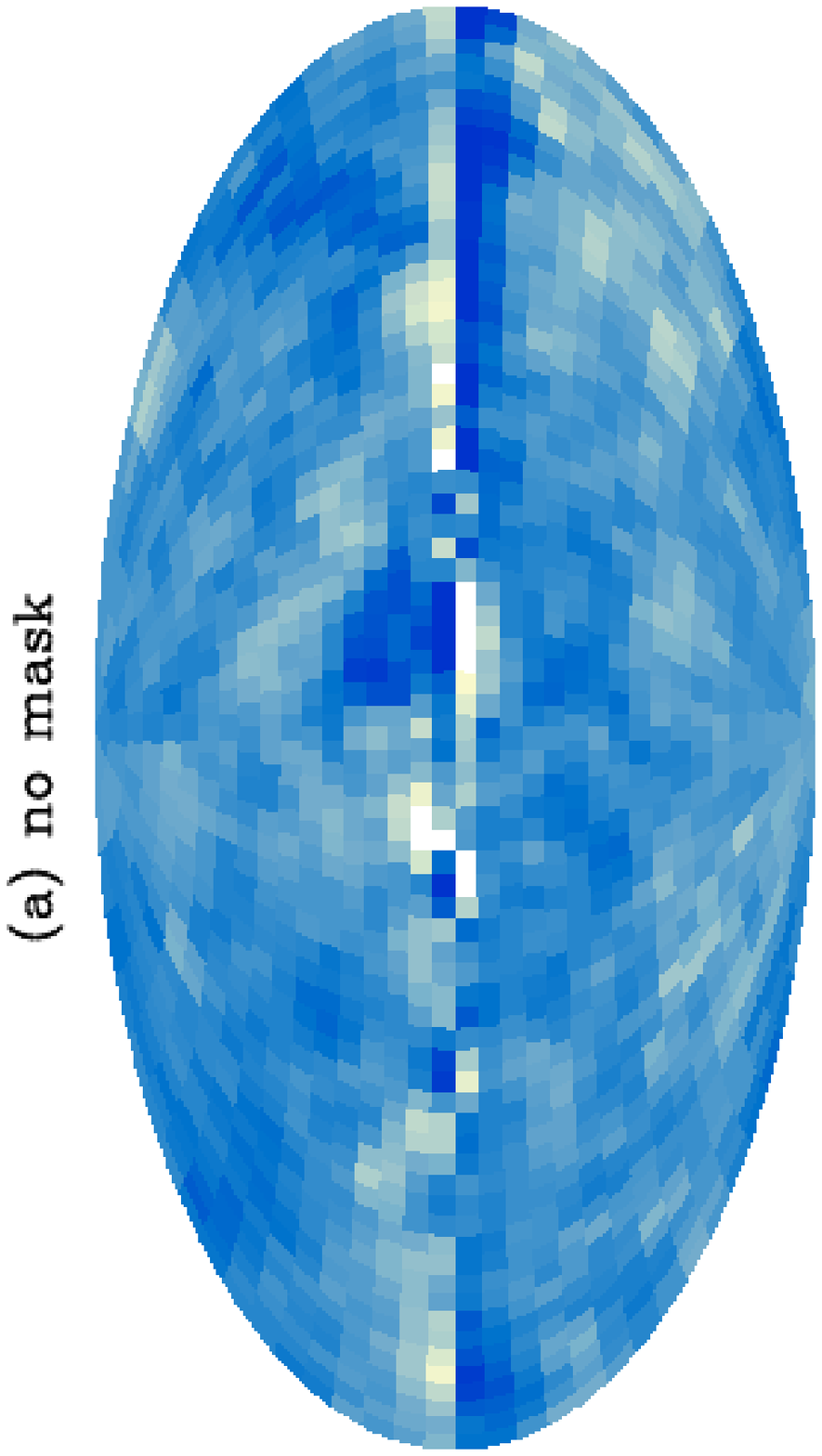}
\includegraphics{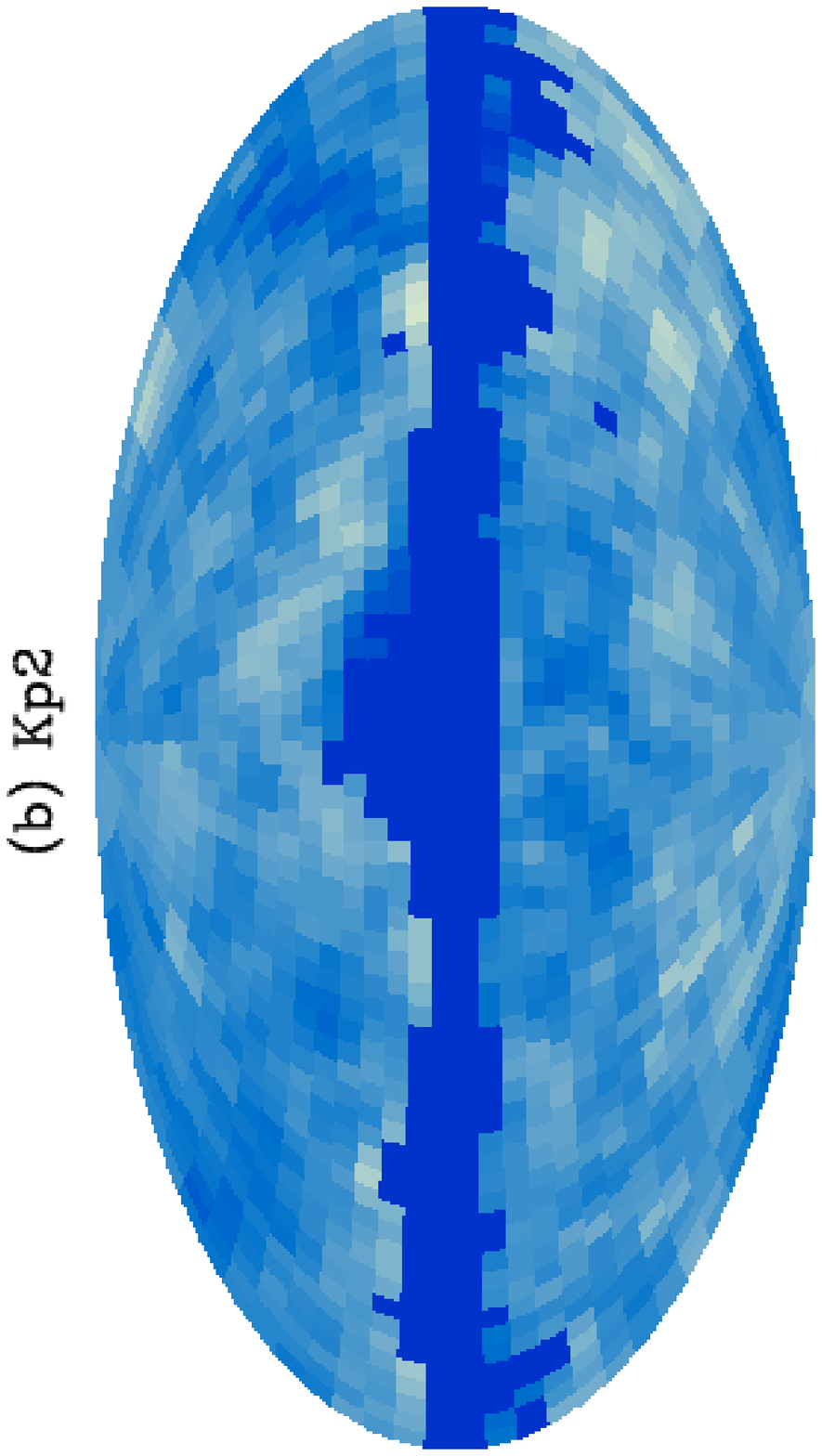}
\includegraphics{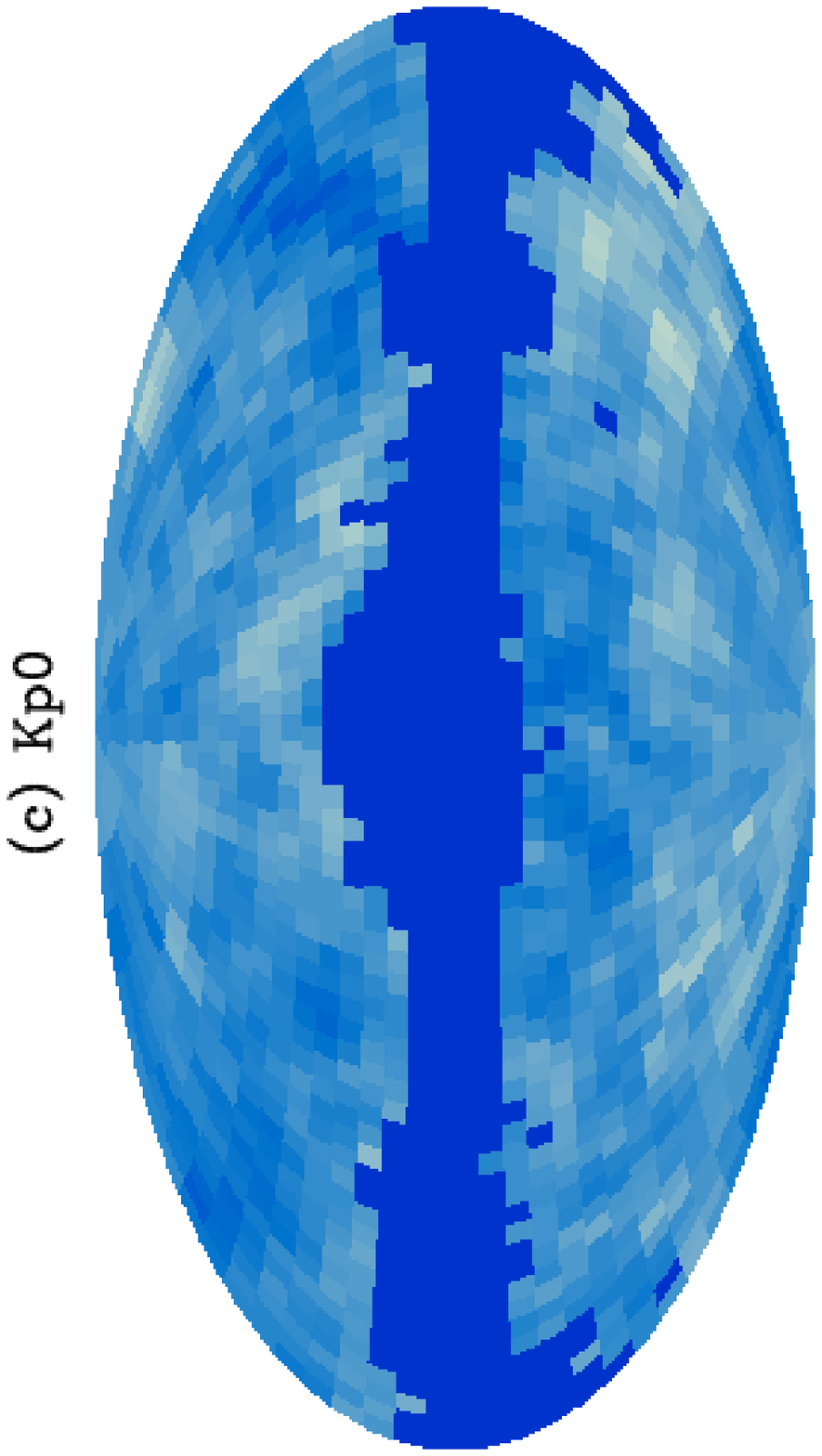}
\includegraphics{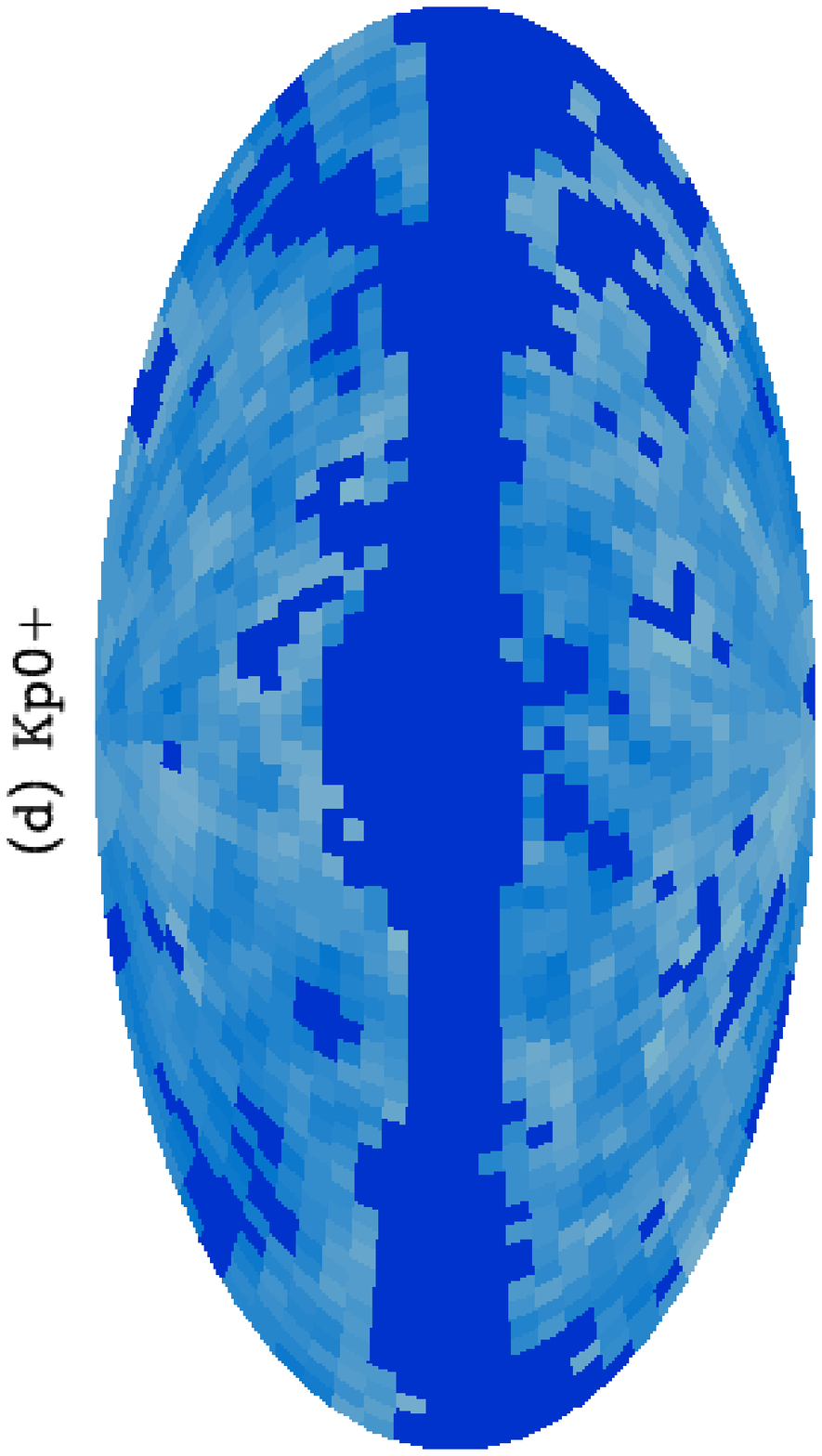}

\caption
{Temperature differences of the low resolution WMAP-ILC and TdOH03 maps. Figure 4a
shows the whole sky, for which the {\it rms} temperature difference is $8.9 \mu K$.
The Kp2 and Kp0 masks are shown in figures 4b and 4c. Figure 4d shows the Kp0+ mask
defined in the text, which consists of the Kp0 mask and those pixels
for which the temperature difference between the two maps differs by more
than $\Delta T/T=3 \times 10^{-6}$  ($8.2 \; \mu K$).}

\label{figure4}

\end{figure*}

\section{Application to WMAP}

\subsection{Low resolution input maps and masks}

The high resolution maps of Figure 1 were first degraded to a pixel
size of $\theta_c = 5^\circ$ as follows. The harmonic coefficients
$a_{\ell m}$ for the maps were computed by applying a spherical
harmonic transform. The low resolution maps were then synthesised from
the $a_{\ell m}$ (discarding the $\ell=0$ and $\ell=1$ multipoles)
after multiplying by a Gaussian beam function of $7^\circ$
FWHM.  The low resolution maps are therefore
equivalent in pixel size and resolution to the simulations described
in the previous Section. The reduction in resolution is required so
that the QML estimator, which requires of ${\cal O}(N_{\rm pix})^3$
operations to evaluate, can be computed quickly. The
$\theta_c=5^\circ$ igloo pixelized maps contain $N_{\rm pix} = 1632$
pixels,

The temperature differences of the low resolution WMAP-ILC and TdOH03
maps are plotted in Figure 4a. Differences are clearly apparent
towards the Galactic plane and in the region of the Ophiucus complex
({\it cf} Figure 4 of Bennett \etals 2003a).  Despite the obvious
differences at low Galactic latitudes, the {\it rms} temperature
differences between these two maps is only $8.9 \;\mu K$.  For
comparison the {\it rms} temperature anisotropy of the WMAP-ILC map at
this resolution is $40.3\; \mu K$. The Kp2 and Kp0 masks are shown in
Figures 4b and 4c. Most of the discrepant pixels in Figure 4a are
eliminated by the Kp2 mask. Even with the Kp0 mask, however, there are
residual differences between the two maps (the {\it rms} temperature
differences for pixels outside the Kp0 mask is $5.6 \; \mu K$, see Table
2). Figure 4d shows a more extensive mask
(denoted Kp0+) which consists of the Kp0 mask and those pixels in the 
difference map that differ by more than $8.2 \;
\mu K$ ($\Delta T/T=3 \times 10^{-6}$). Some of the additional masked pixels, 
for example, in the region of Orion, may reflect inaccuracies in Galactic
subtraction. However, most of the differences at high Galactic
latitude correlate with regions of high emission in the component
separated maps (Figure 1). These differences are most likely
caused by small differences in the way that the two component
separated maps were
constructed (for example, differences in smoothing of the individual
frequency maps) rather than by low level Galactic emission.
Nevertheless, it is interesting to investigate the effects on the
low CMB multipoles by applying the Kp0+ mask.

\subsection{Amplitudes of the Low CMB Multipoles}

The power spectra, up to a multipole of $\ell = 20$, for the four
masks of Figure 4, are plotted in Figures 5 and 6. Figure 5 shows the
results for the WMAP-ILC map and Figure 6 shows the results for the
TdOH3 map. In each plot the solid lines show the results from the PCL
estimator (equation \ref{PCL2}) and the filled circles show the
results of the QML estimator (equation \ref{ML4}). The error bars
on the QML points are computed from the diagonal components of
the covariance matrix (equation (\ref{ML5}) assuming the fiducial $\Lambda$CDM
model).

 In the case of no mask the QML and PCL estimates are identical, as
expected. However, the estimators differ when applied to the masked
maps. Given the discussion in Section 2.4, we would expect the QML
estimates to remain relatively stable as more of the sky is
masked. The PCL estimator is expected to show larger changes as more
of the sky is masked because of the increasing importance of
estimator induced variance. This is exactly what is seen in Figures 5
and 6.

\begin{figure*}
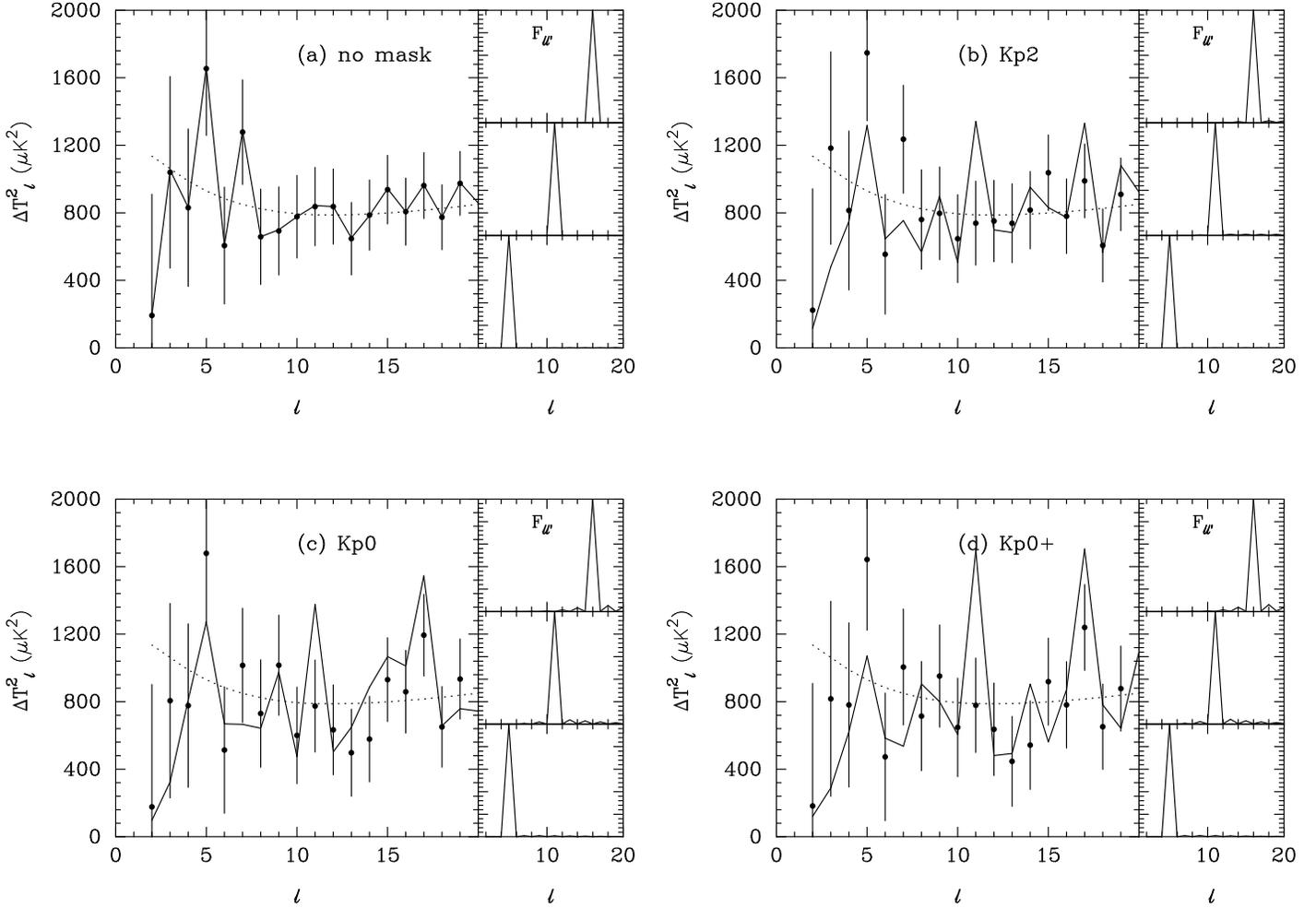


\vskip 5.5 truein

\includegraphics{pgfig2a.ps}
\includegraphics{pgfig2b.ps}
\includegraphics{pgfig2c.ps}
\includegraphics{pgfig2d.ps}

\caption
{The low CMB multipoles computed from the WMAP-ILC map. The filled
circles show the QML estimates of $\Delta T^2_\ell$ and the solid
lines show the PCL estimates. The error bars on the points are
computed from the diagonal components of the Fisher matrix $F_{\ell
\ell^\prime}$, which assumes the fiducial $\Lambda$CDM model. The
panels to the right show three rows of the Fisher matrix. Figure 5a
shows the results if no Galactic mask is imposed. The Kp2 Galactic
mask is used for Figure 5b, the Kp0 mask for Figure 5c, and the Kp0+
mask for Figure 5d. The dotted lines show the power spectrum of the
fiducial $\Lambda$CDM model.}

\label{figure5}

\end{figure*}

\begin{figure*}
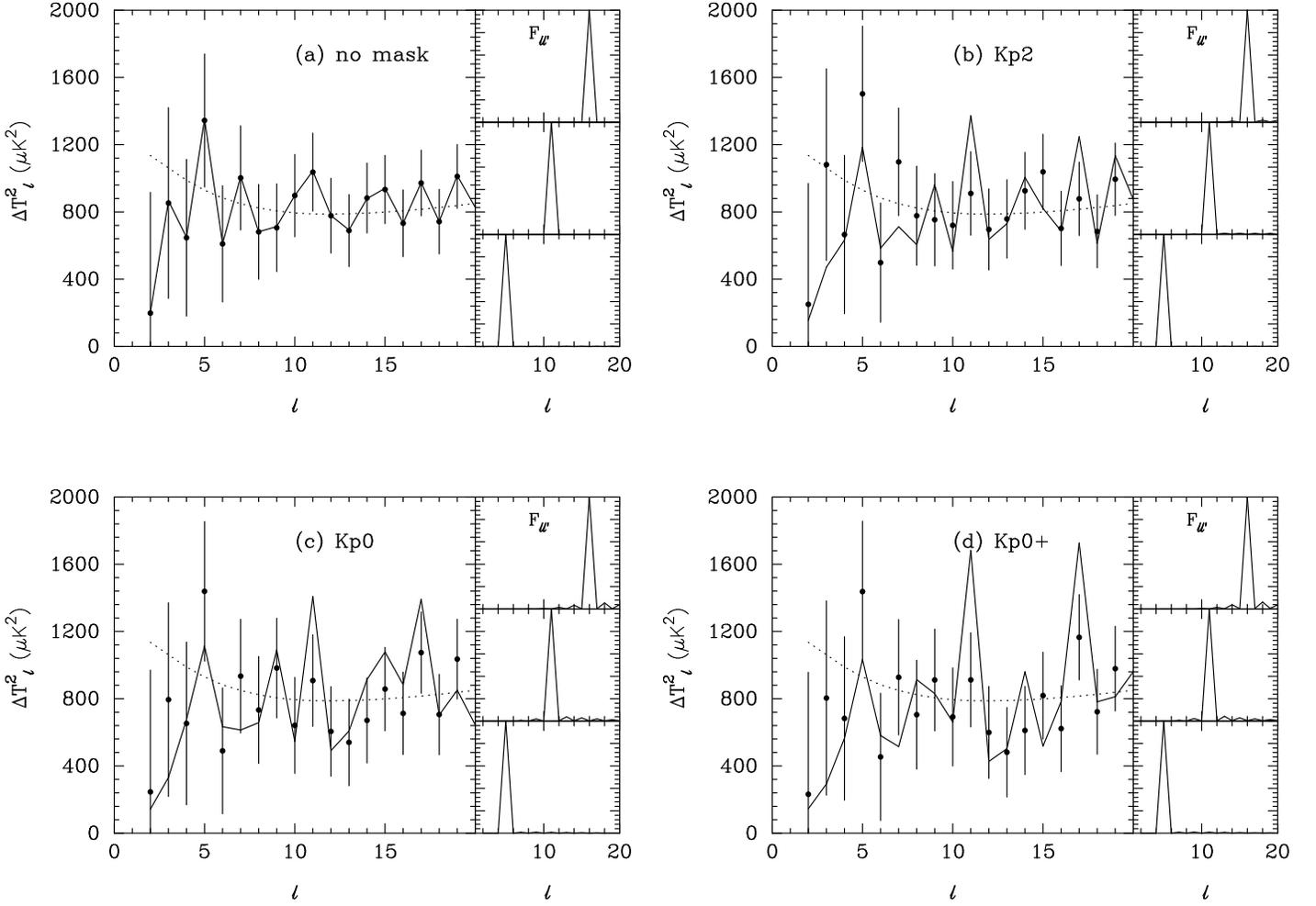


\vskip 5.5 truein

\includegraphics{pgfig3a.ps}
\includegraphics{pgfig3b.ps}
\includegraphics{pgfig3c.ps}
\includegraphics{pgfig3d.ps}

\caption
{As Figure 5, but for the component separated map of TdOH03.  }

\label{figure6}

\end{figure*}

The amplitudes of the quadrupole and octopole for the four masks are
listed in Table 2. For reference, the amplitudes listed in the WMAP
public data release (see also Table 3) are $\Delta T^2_2 = 123 \; (\mu
K)^2$ and $\Delta T^2_3 = 611 \; (\mu K)^2$. These numbers were
determined by applying a PCL estimate to the V and W band maps
 with the Kp2 mask (Hinshaw \etals 2003) and are quite close the PCL
estimates for the WMAP-ILC map listed in Table 2 ($\Delta T^2_2 = 116\;
(\mu K)^2$, $\Delta T^2_3 = 480 \; (\mu K)^2$).  The quadrupole and
octopole amplitudes in the case of no mask are almost identical with
the PCL amplitudes estimated by TdOH03 from the two component
separated maps.  For both maps, the QML quadrupole amplitude is
relatively stable as more of the sky is masked. For example, for the
WMAP-ILC map the QML quadrupole amplitude varies from $192 \;(\mu K)^2$
for no mask to $176 \; (\mu K)^2$ for the Kp0 mask. In contrast, the PCL
quadrupole amplitude halves from $194 \; (\mu K)^2$ for no mask to $97 \;
(\mu K)^2$ for the Kp0 mask. The discussion in Section 2.4 shows that
differences of this order are consistent with the PCL estimator
induced variance. {\it The QML estimates, which are stable for
different masks, provide  more reliable estimates.}

Similar remarks apply to the analysis of the TdOH03 map. The differences
between the QML estimates for the WMAP-ILC and TdOH03 maps provides some
indication of the effects of inaccurate subtraction of Galactic emission.
For the Kp2 mask, the difference in quadrupole amplitudes is $27 (\mu K)^2$
and this rises to $69 (\mu K)^2$ for the Kp0 mask. These differences are broadly
compatible with the estimate given in Bennett \etals (2003a) that the
95\% uncertainty in the quadrupole amplitude caused by modelling
Galactic foregrounds is about $70 \; (\mu K)^2$. Notice that applying the
QML estimates for the more conservative Kp0+ mask are almost identical
to the estimates for the Kp0 mask ({\it cf} Figures 5 and 6). (This is
not surprising because most of the additional masked pixels in the Kp0+
mask are not contiguous with the Kp0 mask.)

Which of these estimates is likely to be the most reliable? The
simulations in Section 2.4 suggest that the results for the Kp2 mask
are likely to be the most accurate. Clearly, the largest differences
between the WMAP-ILC and TdOH3 maps lie within the Kp2 mask, and since
we have demonstrated that the estimator induced variance with the Kp2
mask is small, the Kp2 QML estimates are more likely to be accurate
than the estimates from the unmasked maps. The differences between the
Kp2 estimates for the two component subtracted maps then provide some
indications of the effects of inaccuracies in subtracting Galactic
emission. The QML estimator induced variance does begin to become
important for the Kp0 mask. The small differences between the QML
results for the Kp2 and Kp0 masks are more likely to be caused by
estimator induced variance than to errors in Galactic subtraction
({\it cf} Figures 2 and 3). 

The QML estimates for the first $20$ CMB multipoles with the Kp2 mask
are listed in Table 3. The PCL amplitudes estimated by the WMAP team
are also listed in Table 3 for comparison.  The errors on the QML
estimates are computed from the diagonal components of the covariance
matrix assuming the fiducial $\Lambda$CDM model. Since the estimator
induced variance for the Kp2 mask is small, the covariance matrix is
very accurately diagonal (see the plots of the Fisher matrix in
Figures 5 and 6), and the error estimates are very close to the cosmic
variance errors (equation \ref{I1}). The errors on the WMAP estimates
are computed from equation (13) of Verde \etals (2003) assuming the
fiducial $\Lambda$CDM model.

The difference between the two sets of QML estimates in Table 3
provide some indications of the effects of inaccuracies in subtraction
of Galactic emission. The {\it rms} difference in the estimates from
$\ell=10$ to $\ell =20$ is $88 (\mu K)^2$.  This is compatible with
the $5-10\%$ variations at low multipoles seen in the cross power
spectra of the external template foreground subtracted Q, V and W maps (see Figure 3 of
Hinshaw \etals 2003). In agreement with the WMAP team's conclusions,
these tests suggest that inaccuracies in foreground subtraction are
substantially smaller than the cosmic variance at low multipoles. They
are not, however, negligible and should be included in detailed analyses
({\it e.g.} assessing the statistical significance of the quadrupole
discrepancy, {\it cf} TdOH03, E03b).

\begin{table}
\bigskip

\centerline{\bf   Table 3:  Comparison of QML estimates and errors for the }
\centerline{\bf    Kp2 mask with published WMAP estimates and errors}

\begin{center}

\begin{tabular}{cccc} \hline \hline
\smallskip 
$\ell$  & $\Delta T^2_\ell$ & $\Delta T^2_\ell$ & $\Delta T^2_\ell$ \cr
   &  WMAP (PCL) & TdOH03 (QML) & WMAP-ICL (QML) \cr
\smallskip \cr
$2$ & $123 \pm 845$ & $250$ & $223 \pm 720$  \cr
$3$ & $611 \pm 667$ & $1081$ & $1183 \pm 570$  \cr
$4$ & $756 \pm 548$ & $665$ & $813 \pm 471$  \cr
$5$ & $1257 \pm 466$ & $1502$ & $1748 \pm 402$  \cr
$6$ & $696 \pm 407$ & $498$ & $554 \pm 355$  \cr
$7$ & $830 \pm 364$ & $1097$ & $1236 \pm 319$  \cr
$8$ & $627 \pm 332$ & $777$ & $760 \pm 294$  \cr
$9$ & $815 \pm 307$ & $753$ & $797 \pm 274$  \cr
$10$ & $617 \pm 288$ & $720$ & $647 \pm 260$  \cr
$11$ & $1251 \pm 273$ & $910$ & $739 \pm 248$  \cr
$12$ & $759 \pm 261$ & $696$ & $751 \pm 241$  \cr
$13$ & $714 \pm 252$ & $758$ & $738 \pm 233$  \cr
$14$ & $907 \pm 245$ & $925$ & $816 \pm 229$  \cr
$15$ & $871 \pm 238$ & $1038$ & $1037 \pm 224$  \cr
$16$ & $628 \pm 234$ & $701$ & $780 \pm 221$  \cr
$17$ & $1042 \pm 229$ & $878$ & $989 \pm 218$  \cr
$18$ & $742 \pm 226$ & $684$ & $607 \pm 217$  \cr
$19$ & $947 \pm 223$ & $994$ & $909 \pm 215$  \cr
$20$ & $870 \pm 221$ & $715$ & $782 \pm 214$  \cr
\hline
\end{tabular}
\end{center}

\smallskip

\begin{quote}
{\it Notes to Table 3:} The second column lists the amplitudes of the
low order CMB multipoles from the WMAP public data release, which uses
a PCL estimator and the Kp2 mask. The errors on these numbers are
given by equation (13) of Verde \etals (2003) assuming the fiducial
$\Lambda$CDM model.  The second and third columns list the QML
estimates for the TdOH03 and WMAP-ILC maps with the Kp2 mask
applied. The errors on the QML estimates (which are identical for both
maps) are computed from the diagonal components of the covariance
matrix (equation (\ref{ML5}) assuming the fiducial $\Lambda$CDM model).
\end{quote}

\smallskip

\end{table}

\subsection{Implications for the concordance $\Lambda$CDM cosmology.}

The results from the QML estimator presented in the preceeding
sub-section shows that WMAP estimate of the quadrupole amplitude of
$123 \; (\mu K)^2$ is an underestimate of the true value, which is
more likely to be about $200 \; (\mu K)^2$. Table 4 lists the
probability $P$ of finding a quadrupole amplitude smaller than the
observed value in the fiducial $\Lambda$CDM model. The first row of
Table 4 lists $P$ for the WMAP PCL estimates of Hinshaw \etals
(2003). This was computed by applying the PCL estimator to simulated
CMB skies with the Kp2 mask and so includes estimator induced
variance.  The remaining entries list the probabilities for the QML
estimates computed from equation (3) of E03b ({\it i.e.} assuming a
$\chi^2$ distribution with cosmic variance). These are consistently
higher than the value of $P$ for the PCL estimate.  There is no
discrepancy here. Since the PCL estimator is inferior to the QML
estimator, the frequentist probability $P$ computed for the PCL
estimator from a single realization of the sky will be less reliable
than the frequentist probability determined from the QML estimator.
The results of Table 4 are in agreement with those of TdOH03 and
E03b. If the concordance $\Lambda$CDM model is correct, the
probability $P$ is in the region of 2.5\% -- 4.5\% depending on which
component separated map is used. Similar remarks apply to the QML
octopole amplitudes given in Table 2. The analysis described here
reinforces the conclusion of a previous paper (E03b) that the
discrepancy with the concordance $\Lambda$CDM model at low multipoles
is of order a few percent rather than at the $0.15\%$ level argued by
Spergel \etals (2003).

\begin{table}
\bigskip

\centerline{\bf   Table 4:  Significance of quadrupole discrepancy}

\begin{center} 
\begin{tabular}{ccccc} \hline \hline
method & map & mask & $\Delta T^2_2$ & P \cr
       &     &      &                &   \cr
PCL    & (WMAP analysis)   & Kp2 & 123 & 1.3\% \cr
QML    & WMAP-ILC    & 0 & 192 & 2.6\% \cr
QML    & TdOH03    & 0 & 198 & 2.8\% \cr
QML    & WMAP-ILC    & Kp2 & 223 & 3.6\% \cr
QML    & TdOH03    & Kp2 & 250 & 4.5\% \cr
QML    & WMAP-ILC    & Kp0 & 176 & 2.1\% \cr
QML    & TdOH03    & Kp0 & 245 & 4.4\% \cr
\hline
\end{tabular}
\end{center}

\end{table}

\section{Angular correlation function and the S statistic}

SO3 quantified the lack of structure in the CMB sky on  large angular scales
by evaluating the  statistic  
\begin{equation}
S = \int_{-1}^{1/2} \left [ C(\theta) \right ]^2 \; d\cos\theta, \label{S1}
\end{equation}
where $C(\theta)$ is the angular correlation function computed from the
PCL estimates of the power spectrum $C_\ell$,
\begin{equation}
C(\theta) = {1 \over 4 \pi} \sum (2 \ell + 1) C_\ell P_\ell(\cos\theta). \label{S2}
\end{equation}
By comparing values of $S$ determined from a large number of simulations generated
from the posterior distribution of the $\Lambda$CDM cosmology (using the same sky cut
and estimator as for the real data) they concluded that the probability of
finding a value of $S$ smaller than that observed is about $0.15\%$. 

 The integration limits for the $S$ statistic ($\theta > 60^\circ$)
were chosen a posteriori to correspond to the angular scales over
which the measured $C(\theta)$ is close to zero. Over these angular scales
the quadrupole and octopole make significant contributions to the sum in
(\ref{S2}) but the next few multipoles also make some contribution. The statistic
$S$ is thus not equivalent to a test based on the amplitudes of the quadrupole
and octopole anisotropies alone. It is therefore interesting to analyse the
$S$ statistic using an optimal estimate of the power spectrum in equation (\ref{S2})
and to contrast this analysis with the results of Table 4.

Evidently, the $S$ statistic will show estimator induced variance since it is
 constructed from estimates of the power spectrum. As with the power
 spectrum estimates discussed in Section 2.2, the estimator induced
 variance of the $S$ statistic applied to a cut sky will be higher if
 a PCL estimate of $C_\ell$ is used in the sum (\ref{S2}) than if a
 QML estimate is used.  This is illustrated in Figure 7 which compares
 input and output values of the $S$ statistic determined from
 simulated skies with the Kp2 cut applied using PCL (Figure 7a) and
 QML (Figure 7b) estimates of the power spectrum.
 estimates)\footnote{Note that a direct determination of $C(\theta)$
 by summing over pixels on a cut sky (say as in Gazta\~naga \etal, 
 2003) will have similar estimator induced variance as the summation
 (\ref{S2}) using a PCL estimator.}.

\begin{figure*}
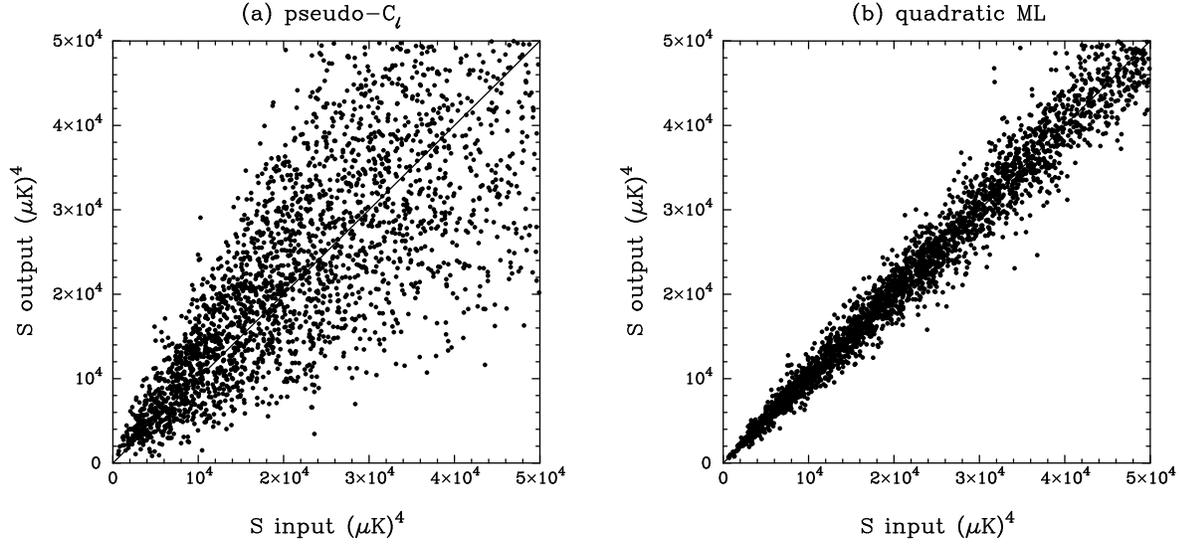


\vskip 3.0 truein

\includegraphics{Sfiga.ps}
\includegraphics{Sfigb.ps}
\caption
{Comparison of the PCL and QML estimates of $S$ statistic (equation
\ref{S1}) from simulations with the Kp2 mask. The abscissae list the
values of the $S$ statistic for the input simulated skies. The
ordinates give the output values from the PCL estimator (Fig 7a) and
QML estimator (Fig 7b) after the application of the Kp2 mask.}

\label{figure7}

\end{figure*}

\begin{figure*}

\vskip 3.0 truein

\includegraphics{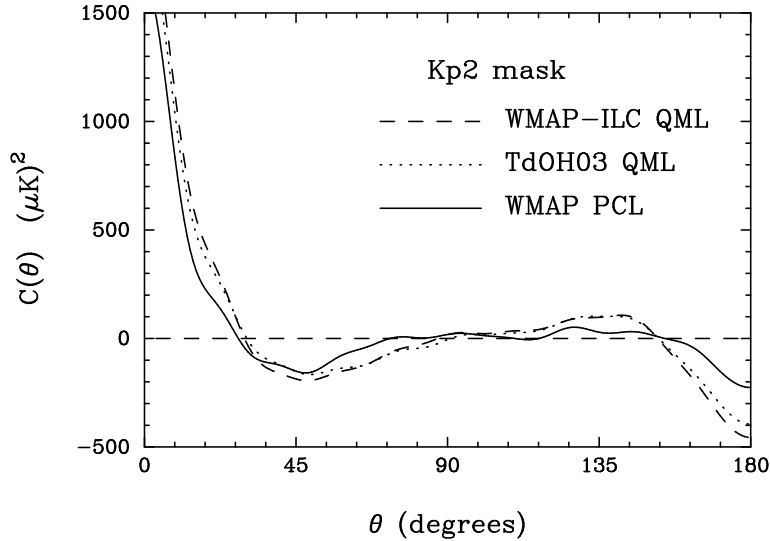}

\caption
{Estimates of the angular correlation function}

\label{figure8}

\end{figure*}

\begin{table}
\bigskip

\centerline{\bf   Table 5:  Significance of the S statistic}

\begin{center} 
\begin{tabular}{ccccc} \hline \hline
method & map & mask & $S$ & $P(S<S_{\Lambda{\rm CDM}})$ \cr
       &     &      &                &   \cr
PCL    & (WMAP analysis)   & Kp2 & 1644. & 0.18\% \cr
QML    & WMAP-ILC    & 0 & 7891 &  6.9\% \cr
QML    & TdOH03    &  0 &  4995 & 3.2\% \cr
QML    & WMAP-ILC    & Kp2 & 11437 & 12.5\% \cr
QML    & TdOH03    & Kp2 &  9929 & 9.7\% \cr
QML    & WMAP-ILC    & Kp0 & 5365 & 3.9\% \cr
QML    & TdOH03    & Kp0 & 5108 & 3.7\% \cr
\hline
\end{tabular}
\end{center}
\begin{quote}
{\it Notes to Table 5:} A comparison of values of the $S$ statistic
(equation \ref{S1}) and its statistical significance.  The first row
gives the results for the publically released WMAP PCL estimates
listed in Table 3. The remaining rows list results using QML estimates
of the power spectrum for various sky cuts. The fourth column list the
value of $S$ (in units of $(\mu K)^5$, The fifth column lists the
frequency $P(S <S_{\Lambda CDM})$ of finding a value of $S$ smaller
than that observered if the fiducial $\Lambda$CDM model is correct.
\end{quote}

\end{table}

For any given Galactic cut, a PCL-based $S$ statistic will {\it
always} be inferior to a QML-based $S$ statistic. An analysis of a
QML-based $S$ statistic must therefore give a more accurate estimate
of the statistical significance of the lack of structure in the CMB
sky than the analysis described by S03.  Figure 8 shows the angular
correlation function using the three power spectrum estimates listed
in Table 3 evaluated with the Kp2 mask (WMAP PCL, TdOH03 QML and
WMAP-ILC QML). Although all three estimates are close to zero on
angular scales $90^\circ \simlt \theta \simlt 150^\circ$, the values
of the $S$ statistic differ significantly, as does the inferred
statistical significance of a discrepancy with the fiducial
$\Lambda$CDM model.  Table 5 lists the values of $S$ (in units of
$(\mu K)^4$) inferred from the publically released PCL power spectrum
estimates and also for QML estimates for various sky cuts. The final
column of Table 5 lists the frequency $P(S< S_{\Lambda\rm{CDM}})$ of
finding a value smaller than the one observed if the fiducial
$\Lambda$CDM model is correct. These frequencies were estimated
directly from the simulated CMB skies, using identical estimators and
sky cuts to those applied to the real data. All of the QML estimates
give frequencies of a few percent, in qualitative agreement with the
frequencies listed in Table 4.

Only the first row of Table 5, using the WMAP PCL estimates gives a
low frequency of $0.18\%$. This low frequency is consistent with the
analysis of SO3.  It is also broadly consisent with the low frequency
of $0.35\%$ of finding a quadrupole {\it and} octopole amplitude
lower than the WMAP PCL estimates given the fiducial $\Lambda$CDM
model (E03b). However, since these frequencies are based on
PCL estimates, they {\it must necessarily} be less reliable that the
frequencies based on QML power spectrum estimates. In particular,
the low frequency inferred by SO3 from the $S$ statistic is simply
an unfortunate consequence of the way that PCL  estimator induced variance
has affected  estimates of the low CMB multipoles. The more
reliable QML (or QML-based)  estimates listed in Tables 4 and 5
show that a more accurate estimate of the significance level of
the discrepancy with the fiducial $\Lambda$CDM model is of order
a few percent.

\section{Conclusions}

In this paper, we have investigated the effects of sky cuts on PCL and
QML estimators using numerical simulations. For QML estimators, the
estimator induced variance of the quadrupole amplitude is less than
$40 \; (\mu K)^2$ for the WMAP Kp0 mask and, for most purposes, is
negligible when the less severe Kp2 mask is applied. In contrast, at
low multipoles the PCL estimator begins to break down for the Kp0
mask, since the estimator induced dispersion for such a large sky cut
is comparable to the signal. A QML estimator is therefore preferable
to a PCL estimator, and for small enough sky cuts is capable to
returning almost the exact amplitudes of the low multipoles for our
realization of the CMB sky.

The PCL and QML estimators have been applied to the Galaxy subtracted
maps produced by B03b and TdOH03 to estimate the amplitudes of the CMB
multipoles at $\ell \le 20$.  The QML estimates (in contrast to the
PCL estimates) are found to be stable to the imposed Galactic
cut. This stability, and the agreement between the power spectra from
the two maps, suggests that inaccuracies in Galactic subtraction
introduce errors of order $10\%$ or less in the amplitudes of the low
multipoles.

The QML quadrupole and octopole amplitudes are found to lie in the
ranges $\Delta T_2^2 = 176 - 250\; (\mu K)^2$ and $\Delta T_3^2 = 794
- 1183\; (\mu K)^2$ and are more likely to lie at the upper ends of
these ranges since these values correspond to the Kp2 Galactic cut,
for which the estimator induced variance and Galactic emission is
small. In contrast, the WMAP team derived values $\Delta T_2^2 = 123
\; (\mu K)^2$ and $\Delta T_3^2 = 611 \;(\mu K)^2$ by applying a PCL
estimator to maps with the Kp2 sky cut. There can be no question that
the QML estimates are more reliable than the PCL estimates. There is,
therefore, strong evidence that the discrepancy between the quadrupole
and octopole amplitudes and those expected in the concordance
$\Lambda$CDM model is considerably less significant than the $0.15\%$
estimated by SO3. This is consistent with the analysis of the $S$
statistic described in Section 4 using QML estimates.  The results
summarized in Tables 4 and 5, in fact suggest, a significance level of 
the low multipole discrepancy of a few
percent.

The results described here are compatible with those of TdOH03. These
authors derived quadrupole and octopole amplitudes of $\Delta T_2^2 =
202 \; (\mu K)^2$ and $\Delta T_3^2 = 856 \; (\mu K)^2$ (very close to
the numbers given in Table 2) from an analysis of their all sky
component separated map. They argued that the residual contamination
from inaccurate Galactic subtraction was small enough, and confined
to a sufficiently small number of pixels close to the Galactic plane,
that the results of the all-sky analysis should give accurate estimates
of the quadrupole and octopole amplitudes. This is consistent the
analysis presented in this paper, since the QML estimates are 
found to be insensitive to the Galactic cut. 

The results presented here weaken the case that any exotic new physics
is required to explain the amplitudes of the low CMB multipoles.

\medskip

\noindent
{\bf Acknowledgments:} I thank Max Tegmark for supplying copies of
the TdOH03 maps and to various members of the Planck analysis group
at Cambridge for useful discussions.

\end{document}